\documentclass[intlimits,twoside,a4paper]{article}

\usepackage[cp1251]{inputenc}
\usepackage[eqsecnum]{cmpj3}

\usepackage{bm}

\issue{2018}{21}{4}{43501}
\doinumber{10.5488/CMP.21.43501}

\title[Cluster expansion for the description of condensed state: crystalline cell approach
]%
{Cluster expansion for the description of condensed state: crystalline cell approach%
}
\author[G.S.~Bokun, M.F.~Holovko]{G.S.~Bokun\refaddr{label1}, M.F.~Holovko\refaddr{label2}}
\addresses{
	\addr{label1}Belarusian  State Technological University, 13a Sverdlov St., 220006 Minsk, Belarus
	\addr{label2} Institute for Condensed Matter Physics of the National Academy of Sciences of Ukraine, \\
	1 Svientsitskii St., 79011 Lviv, Ukraine
}

\date{Received May 2, 2018,  in final form July 10, 2018}

\begin{document}

\maketitle

\begin{abstract}
A well-known cluster expansion, which leads to virial expansion for the free
energy of low density systems, is modified in such a way that it becomes
applicable to the description of condensed state of matter. To this end, the
averaging of individual clusters over the states of an ideal gas is replaced
by the averaging over the states of a non-correlated crystal using
single-particle cell potentials. As a result, we arrive at the expansion of
the partition function in correlations on the basis of single-particle
functions corresponding to the multiplicative approximation. The cell
potentials defining these functions are found from the condition of the
minimum of the remainder in the constructed decomposition. 
\keywords lattice models, cluster expansions, single-particle cell potential, free energy
\pacs 5.20.-y, 61.72jd, 64.30.-t, 65.40.-b
\end{abstract}


\section{Introduction}

The cellular theory, which is based on the structuring of the states of the
system, has made it possible to solve a number of important problems in the
physics of condensed state \cite{Prig57,Fren45}. The lattice version of this
theory has been used to address fundamental problems of statistical physics
as well as to calculate specific properties of various systems
\cite{Yukhnovskii87,Bokun18}. The development and introduction of
nanomaterials has driven these approaches to be applied to the description of
highly heterogeneous structures, which, in addition to large gradients of
order parameter fields, are characterized by a heterogeneous state
accompanied by phase transitions of various types \cite{Bisq04,Ciach82}.
Here, as in the case of homogeneous systems, it seems effective to involve
mean-field representations based on the use of cell potentials
\cite{Narkevich,diCaprio09}.

The present paper is devoted to the development of the basis of this
approach. To this end, a system of particles in the mean field of
single-particle cell potentials is used as the reference system. The cell
potentials determine the average forces acting on a particle fixed in a
selected cell from the particles distributed in other molecular cells.
Initially, only those states are taken into consideration where each cell of the system is occupied by one
molecule or is vacant.
The heterogeneity in the system is taken into account
by the fact that the cell volumes are not the same.
In order to calculate the partition function of the initial condensed system,
the perturbation theory is used based on the expansions over the states of
the reference system using generalized Mayer functions. These functions
contain not only intermolecular interaction potentials but also the necessary
mean potentials. The latter are determined from the condition for optimizing
the deviation of the properties of the reference and the real systems.
Furthermore, the thermodynamic consistency of the theory is studied at the
level of calculation of the first derivatives with respect to thermodynamic
parameters. The developed microscopic approach is generalized to the case of
long-range interaction. A possibility to take account of a more complete set of
occupation number values is discussed.

The paper is arranged as follows. In section~\ref{sec2}, the lattice model and the description of a corresponding
reference system are presented. In order to take into account the inter-particle correlations in
section~\ref{sec3}, we formulate the perturbation theory based on the expansions over the states of the
reference system. In section~\ref{sec4}, we formulate the optimal choice of a single-particle cell potentials needed
for the description of the reference system. In section~\ref{sec5}, the verification of thermodynamic self-consistency of the presented approach is considered at the level of calculation of the first derivatives
with respect to thermodynamic parameters. In section~\ref{sec6}, the generalization of the considered approach for
the case of a more complete set of occupation numbers values and the generalization to the case of the
systems with long-range interaction are discussed. We conclude in section~\ref{sec7}.

\section{The model and the reference system}\label{sec2}

In this paper we consider a lattice model with the Hamiltonian
\begin{equation} \label{buk23} H_{M} =\frac{1}{2} \sum \limits _{i=1}^{M}\sum \limits
_{j\left(i\right)}^{Z}\Phi \big(q_{n_{i} } ,q_{n_{j} } \big)  +\sum \limits _{i=1}^{M}
\mu _{i}  n_{i}\,,  \end{equation}
where $M$ is the total number of lattice sites (i.e., cells in the system), $\sum
 _{j\left(i \right)}^{Z} $ denotes consecutive summation over all
nodes surrounding the selected node $i$ taking into account $Z$
counted neighbors, $\mu_{i}$ is the chemical
potential value in the node $i$. The classification of states considered is similar to that adopted in
 \cite{Rott79}. The variable  $q_{n_{i}}$
determines the position of the particle ($n_i=1_i$) or the vacancy
($n_i=0_i$) in the cell $i$, $\Phi \big(q_{n_{i} } ,q_{n_{j} }\big)$ is the interaction potential between two particles
with coordinates $q_{n_{i}}$ and $q_{n_{j}}$. We suppose that this potential is short-ranged enough and can be taken, for instance, in the
Lennard-Jones form \cite{Yukhnovsky80}.

In order to describe the considered model with the Hamiltonian (\ref{buk23}), we apply the concept of the reference system widely used
in the statistical theory of various condensed systems. Usually the reference system is a simplified version of the real condensed system. It should
include
the main features of the real model and should be described analytically with sufficient accuracy. For example, the simplest reference system is the
model
of the ideal gas. The application of this model leads to the virial density expansion for thermodynamic properties of real gases \cite{Kampen61}.
Another important reference system is the model of hard spheres, which has been successfully employed in the modern liquid state theory
\cite{Yukhnovsky80}. In this paper, to study the system with the Hamiltonian (\ref{buk23}), we use the reference system that can be
described by the Hamiltonian
\begin{equation}
 \label{buk1}
 H_{0} =\sum \limits _{i=1}^{M}\mu _{i}  n_{i} +\sum \limits
_{i=1}^{M}\sum \limits _{j\left(i\right)}^{Z}\phi _{j} \left(q_{n_{i} } \right)
 \end{equation}
represented by single-particle
cell potentials $\phi _{j} \left(q_{n_{i}}\right)$, which has the
meaning of the potential of an external field whose source is located
formally at the center of the cell $j$. $\phi _{j} \left(q_{n_{i} } \right)$
depends on the variables that determine the position of particles or vacancies in the cell
$i$ and parametrically is a function of the quantities characterizing the
average distribution of particles or vacancies in the system and its macroscopic state. In
this case, $n = 1$ corresponds to the distribution of particles and $n = 0$ corresponds
to that of vacancies.

The connection between the Hamiltonian $H_M$ and $H_0$ and the calculation of the single-particle potentials is discussed in the next sections.
In this section we consider only the description of the reference system without elaboration of the single-particle cell potential $\phi _{j}
\left(q_{n_{i}}\right)$.

We consider the case of an inhomogeneous system with inhomogeneity
characterized by the field of mean occupation numbers
\begin{equation} \label{buk2}
 \rho _{0} =\langle n_{i} \rangle _{0}\,.
\end{equation}

In order to simplify further expressions, we use abbreviations wherever
possible, for example $U(q_{n_{i} } )=U_{n_{i} } $ denote the potential
acting on a particle or a vacancy in the position $q_{n_{i} }$. In order to take into
account the variable number of particles and vacancies in the system, we consider both
$q_{1_{i} } $ and $q_{0_{i} } $ as two states of a certain ``virtual''
particle. The first one corresponds to the position of a real particle and
the second one to that of a vacancy. In this case, it is possible to replace
fixed particles by arbitrary ones.

This can be seen from the definition of $Z_{M}$
\begin{equation} \label{buk3}
 Z_{M} =\sum \limits _{N=0}^{M}\frac{1}{N!}  Q_{N} \text{e}^{-
\mu N}.
\end{equation}

Due to indiscernibility of particles, $N!$ is reduced, which makes it
possible to represent $Z_{M}^{ (0)} $ of the reference system in the form
\begin{equation}
 \label{buk4}
Z_{M}^{(0)} =\sum \limits _{n_{1}
=0}^{1}\int \limits _{\omega _{1} } \rd q_{n_{1} }   ...\sum \limits _{n_{i} =0}^{1}
\int \limits _{\omega _{i} } \rd q_{n_{i} }   ...\sum \limits _{n_{M} =0}^{1}\int \limits
_{\omega _{M} } \rd q_{n_{M} }    \exp \Bigg\{-\beta \left[\sum \limits _{l=1}^{M}
\left(\mu _{l} n_{l} +U_{n_{l} } \right) \right]\Bigg\}, \end{equation}
where $\beta=1/(kT)$, $k$ is the Boltzmann constant, $T$ is the temperature, $\omega_i$ is the volume of the lattice cell~$i$,
\begin{equation} \label{buk5} U_{n_{l} } =\sum \limits _{j\left(l\right)}^{Z}\phi _{j}
\left(q_{n_{l} } \right). \end{equation}

In accordance with the representation (\ref{buk4}), the particle distribution
function over the volume of the system turns out to be factorized and can be written in
the form
\begin{equation} \label{buk6} D_{M}^{\left(0\right)} =\prod \limits _{i=1}^{M}\rho
\left(q_{n_{i} } \right),
\end{equation}
where
\begin{equation}
\label{buk7}
\rho \left(q_{n_{i} } \right)=\exp \left[-\beta \left(\mu _{i} n_{i} +U_{n_{i}
} \right)\right].
\end{equation}

The normalization of this function, respectively, is represented by the expression

\begin{equation} \label{buk8a}Z_{M}^{
\left(0\right)} =\prod \limits _{i=1}^{M}z_{i}^{0}\,,  \end{equation}
where
\begin{equation} \label{buk8} z_{i}^{0} =\sum \limits _{n_{i} =0}^{1}\int \limits _{
\omega _{i} }\exp   \left[-\beta \left(\mu _{i} n_{i} +U_{n_{i} } \right)\right] \rd q_{n_{i}
}\,,  \end{equation}
\begin{equation} \label{buk9} z_{i}^{0} =Q_{0_{i} } + \text{e}^{\beta \mu _{i} } Q_{1_{i} }\,,  \end{equation}
\begin{equation}
\label{buk10} Q_{n_{i} } =\int \limits _{\omega _{i} }\exp \Bigg[-\beta \sum \limits
_{j\left(i\right)}^{Z}\phi _{j} \left(q_{n_{i} } \right) \Bigg] \rd q_{n_{i} }\,.  \end{equation}

Based on equations (\ref{buk6})--(\ref{buk10}), we write the expression for the
normalized distribution function, which is necessary for averaging the
magnitudes of the singlet, binary, and other types. Thus, denoting the
normalized functions by a cap above the notations of respective functions, we
write
\begin{equation} \label{buk11} \widehat{D}_{M} =\prod \limits _{i=1}^{M}\widehat{\rho
} \left(q_{n_{i} } \right), \end{equation}
where
\begin{equation} \label{buk12} \widehat{\rho }\left(q_{n_{i} } \right)=\frac{\exp \left
[-\beta \left(\mu _{i} n_{i} +U_{n_{i} } \right)\right]}{z_{i}^{0} }\,.  \end{equation}

We transform the above expressions based on the relation between $\mu_{i}$
and the mean values of the occupation numbers $\rho _{1_{i} }$ and $\rho
_{0_{i} }$.
Integrating
(\ref{buk12}) with respect to $q_{n_{i}}$, we find
\begin{equation} \label{buk13} \rho _{n_{i} } =\frac{\text{e}^{\beta \mu _{i} n_{i} }
Q_{n_{i} } }{z_{i}^0 }.  \end{equation}
From (\ref{buk13}) it follows that
\begin{equation} \label{buk14} \frac{\rho _{1_{i} } }{\rho _{0_{i} } } =\frac{\text{e}^{\beta
\mu _{i} } Q_{1_{i} } }{Q_{0_{i} } }.  \end{equation}
The relation (\ref{buk14}) allows us to write
\begin{equation} \label{buk15} \text{e}^{\beta \mu _{i} } =\frac{\rho _{1_{i} } Q_{0_{i} }
}{\rho _{0_{i} } Q_{1_{i} } }.  \end{equation}
Substituting (\ref{buk15}) into (\ref{buk9}), we obtain
\begin{equation} \label{buk16} z_{i}^{0} =Q_{0_{i} } +\frac{\rho _{1_{i} } Q_{0_{i}
} }{\rho _{0_{i} } } =\frac{Q_{0_{i} } }{\rho _{0_{i} } }.  \end{equation}
The substitution of (\ref{buk15}) and (\ref{buk16}) allows us to write
\begin{equation}
\label{buk17} \widehat{\rho }\left(q_{n_{i} } \right)=\rho _{n_{i} } \frac{\exp \left
[-\beta \sum \limits _{j\left(i\right)}^{Z}\phi _{j} \left(q_{n_{i} } \right) \right
]}{Q_{n_{i} } }.  \end{equation}

Let us consider the averaging that employs (\ref{buk11}), for example the
characteristics of a binary type $L\big(q_{n_{i} } ,q_{n_{j} } \big)$
according to the definition
\begin{equation} \label{buk18} {\langle L_{ij} \rangle_{0} =\sum \limits _{n_{1}
=0}^{1}\sum \limits _{n_{2} =0}^{1}...\sum \limits _{n_{m} =0}^{1}\int \limits _{
\omega _{1} }\widehat{\rho }    \left(q_{n_{1} } \right)...}  {\int \limits _{
\omega _{j} }\widehat{\rho } \big(q_{n_{j} } \big)...\int \limits _{\omega _{M}
}\widehat{\rho } \big(q_{n_{j} } \big)...\int \limits _{\omega _{m} }\widehat{
\rho } \left(q_{nm} \right)L\big(q_{n_{i} } q_{n_{j} } \big)}  {}. \end{equation}

Transposing the summation in~(\ref{buk18}) and taking into account the
independence of integration variables and the normalized condition
\begin{equation} \label{buk19} \sum \limits _{n_{i} =0}^{1}\int \limits _{\omega _{i}
}\widehat{\rho }  \left(q_{n_{i} } \right) \rd q_{n_{i} } =\sum \limits _{n_{i} =0}^{1}
\rho _{n_{i} }  =1 \end{equation}
we obtain
\begin{equation} \label{buk19a} \langle L_{ij} \rangle_{0} =\sum \limits _{n_{i=0} }^{1}\sum \limits _{n_{j} =0}^{1}\int \limits
_{\omega _{i} } \rd q_{n_{i} }    \int \limits _{\omega _{j} } \rd q_{n_{j} }  L\big(q_{n_{i}
} ,q_{n_{j} } \big) \widehat{\rho }\left(q_{n_{i} } \right)\widehat{\rho }
\big(q_{n_{j} } \big)\,,\end{equation}
or
\begin{equation} \label{buk20} \langle L_{ij} \rangle_{0} =\sum \limits _{n_{i=0} }^{1}\sum \limits
_{n_{j} =0}^{1}L_{n_{i,} n_{j} }   \rho _{n_{i} } \rho _{n_{j} }\,,  \end{equation}

\begin{equation}
\label{buk21} L_{n_{i,} n_{j} } =\int \limits _{\omega _{i} } \rd q_{n_{i} }  \int \limits
_{\omega _{j} } \rd q_{n_{j} }  L\big(q_{n_{i} } ,q_{n_{j} } \big)\widehat{F}_{11}
\left(q_{n_{i} } \right)\widehat{F}_{11} \big(q_{n_{j} } \big) . \end{equation}

The functions $\widehat{F}_{11} \left(q\right)$ in (\ref{buk21}) correspond
to singlet distribution functions $F_{11}$ in the approximation of the method
of conditional distributions \cite{Rott79}
\begin{equation} \label{buk22} \widehat{F}_{11} \left(q_{n_{i} } \right)=\frac{1}{Q_{n_{i}
} } \exp \Bigg[-\beta \sum \limits _{j\left(i\right)}^{Z}\phi _{j}  \left(q_{n_{i}
} \right)\Bigg]. \end{equation}

\section{Perturbation theory}\label{sec3}

In this section, we consider a perturbation scheme for the treatment of the remainder
\begin{equation}
 \label{buk25} \Delta H_{M} =H_{M} -H_{0}=\frac{1}{2} \sum \limits _{i=1}^{M}\sum
\limits _{j\left(i\right)}^{Z}\Delta \phi \big(q_{n_{i} } ,q_{n_{j} } \big)\,,
 \end{equation}
where
\begin{equation}
\label{buk26}
\Delta \phi \big(q_{n_{i} } ,q_{n_{j} } \big)=\Phi \big(q_{n_{i}
} ,q_{n_{j} } \big)-\phi _{j} \big(q_{n_{i} } \big)-\phi _{i} \big(q_{n_{j}
} \big).
\end{equation}

Now, we represent the partition function of the original system in the form
\begin{equation}
 \label{buk27}
  Z_{M} =Z_{M}^{\left(0\right)} \langle \text e^{-\beta \Delta H_{M}
} \rangle_{0}\,,
 \end{equation}
where $\langle\dots\rangle_{0}$ is the averaging represented by the expression
\begin{equation} \label{buk28} \langle L\rangle_{0} =\sum \limits _{n_{1} =0}^{1}...\sum \limits
_{n_{M} =0}^{1}\int \limits _{\omega _{1} } \rd q_{n_{U} } \widehat{\rho }   \left(q_{n_{U}
} \right)...\int \limits _{\omega _{M} }\widehat{\rho } \left(q_{n_{M} } \right) \rd q_{n_{M}
}  L\,. \end{equation}

To calculate (\ref{buk27}), a cumulant expansion \cite{Kubo62,Yukhnovsky80}
is used, leading to an expansion in powers of the density if the averaging in
(\ref{buk27}) is performed over the states corresponding to an ideal gas
\cite{Kampen61}. In our case, the distribution characteristic of an ideal
crystal is used as the reference system. This allows us to obtain a suitable
description of the properties of a condensed system. The virial coefficients
in this case become density functions. In other words, the transformation
(\ref{buk27}) is an expansion over the cluster correlations, although
formally it has the form of an expansion over the Mayer functions. For the
latter we use the renormalized Mayer functions of the form
\begin{equation} \label{buk29}
f\big(q_{n_{i} } ,q_{n_{j} } \big)=\exp \Big[-\beta
\Delta \phi \big(q_{n_{i} } ,q_{n_{j} } \big)\Big]-1.
\end{equation}

We should note that for vacancies
$\Phi(q_{0_i},q_{0_j})=\Phi(q_{0_i},q_{1_j})=\Phi(q_{1_i},q_{0_j})=0$.
However, the Mayer functions $f(q_{n_i},q_{n_j})\ne 0$ because in this case
$\bigtriangleup\phi(q_{n_i},q_{n_j})\ne 0$.

Using the procedure of group expansion in (\ref{buk27}), we obtain
\begin{align}
\label{buk30}
Z_{M} &=Z_{M}^{\left(0\right)}\exp \Bigg[\frac{1}{2}\sum \limits
_{i=1}^{M}\sum \limits _{j\left(i\right)}^{Z}\sum \limits _{n_{i} =0}^{1}\sum \limits
_{n_{j} =0}^{1}\int \limits _{\omega _{i} } \rd q_{n_{i} }   \int \limits _{\omega _{j}
}\rd q_{n_{j} }  \widehat{\rho }\left(q_{n_{i} } \right)\widehat{\rho }\big(q_{n_{j}
} \big) f\big(q_{n_{i} } ,q_{n_{j} } \big) \nonumber  \\ 
&+\frac{1}{6}\sum \limits
_{i=1}^{M}\sum \limits _{j\left(i\right)}^{Z}\sum \limits _{l\left(i\right)}^{Z}\sum \limits _{n_{i} =0}^{1}\sum \limits
_{n_{j} =0}^{1}\sum \limits
_{n_{l} =0}^{1}\int \limits _{\omega _{i} }\rd q_{n_{i} }   \int \limits _{\omega _{j}
}\rd q_{n_{j} } \int \limits _{\omega _{l}
}\rd q_{n_{l} }  \widehat{\rho }\left(q_{n_{i} } \right)\widehat{\rho }\big(q_{n_{j}
} \big) \widehat{\rho }\left(q_{n_{l}
} \right) \nonumber \\
&\times f\big(q_{n_{i} } ,q_{n_{j} } \big) f\big(q_{n_{j} } ,q_{n_{l} } \big) f\big(q_{n_{i} } ,q_{n_{l} } \big)
+... \Bigg].
\end{align} 

Or, due to representations (\ref{buk20}) and (\ref{buk21}),
\begin{align}
\label{buk31} 
Z_{M}&=Z_{M}^{\left(0\right)}\exp \Bigg[\frac{1}{2} \sum \limits _{i=1}^{M}
\sum \limits _{j\left(i\right)}^{Z}\sum \limits _{n_{i} =0}^{1}\sum \limits _{n_{j}
=0}^{1}\rho _{n_{i} } \rho _{n_{j} } f_{n_{i} n_{j} } \nonumber \\
&+\frac{1}{6}\sum \limits
_{i=1}^{M}\sum \limits _{j\left(i\right)}^{Z}\sum \limits _{l\left(i\right)}^{Z}\sum \limits _{n_{i} =0}^{1}\sum \limits
_{n_{j} =0}^{1}\sum \limits_{n_{l} =0}^{1}\rho_{n_i}\rho_{n_j}\rho_{n_l}f_{n_in_jn_l}+ ...    \Bigg]\,, 
\end{align}
where
\begin{align}
\label{buk32} 
f_{n_{i}n_{j} }&=\int \limits _{\omega _{i} } \rd q_{n_{i} }  \int \limits
_{\omega _{j} } \rd q_{n_{j} }  f\big(q_{n_{i} } ,q_{n_{j} } \big)\widehat{F}_{11}
\big(q_{n_{i} } \big)\widehat{F}_{11} \big(q_{n_{j} } \big)\,,\\
f_{n_{i}n_{j}n_l }&=\int \limits _{\omega _{i} } \rd q_{n_{i} }  \int \limits
_{\omega _{j} } \rd q_{n_{j} } \int \limits _{\omega _{l} } \rd q_{n_{l} }  f\big(q_{n_{i} } ,q_{n_{j} } \big)
f\big(q_{n_{j} } ,q_{n_{l} } \big)f\left(q_{n_{i} } ,q_{n_{l} } \right) \nonumber \\
&\times \widehat{F}_{11}
\left(q_{n_{i} } \right)\widehat{F}_{11} \big(q_{n_{j} }\big) \widehat{F}_{11} \left(q_{n_{l} }\right).
 \end{align}

From (\ref{buk31}), we have the cluster expansion for the free energy
\begin{align} \label{buk33}
F&=-kT\ln Z_{M} =-kT\Bigg[\ln Z_{M}^{0} +\frac{1}{2}\sum \limits _{i=1}^{M}\sum
\limits _{j\left(i\right)}^{Z}\sum \limits _{n_{i} =0}^{1}\sum \limits _{n_{j} =0}^{1}
\rho _{n_{i} } \rho _{n_{j} } f_{n_{i} n_{j} } \nonumber  \\
&+\frac{1}{6}\sum \limits
_{i=1}^{M}\sum \limits _{j\left(i\right)}^{Z}\sum \limits _{l\left(i\right)}^{Z}\sum \limits _{n_{i} =0}^{1}\sum \limits
_{n_{j} =0}^{1}\sum \limits_{n_{l} =0}^{1}\rho_{n_i}\rho_{n_j}\rho_{n_l}f_{n_in_jn_l}+ ...    \Bigg].
\end{align}

\section{An optimal choice of a single-particle potential}\label{sec4}

In order to apply the considered theory, we should specify the
single particle potential. In this section we propose an optimal choice for this potential. This approach is in some sense similar to the
problem
 of the connection between models with soft and hard core repulsions \cite{Andersen} which are successfully used in modern liquid state theory
 \cite{Barker76}.
The equation defining the single-particle potentials of the reference system
is determined from the self-consistent condition, which has several different
formulations. One of them is connected with the extremum of the remainder in
the expansion (\ref{buk30}) due to the fact that the sum of all terms
contained in (\ref{buk30}) does not depend on the choice of single-particle
potentials. Since a change of the potentials leads to redistribution of
contributions of individual terms, the best choice would be the one with the
maximum contribution of the terms that are taken into account. This is
analogous to the requirement of the minimum susceptibility of the system to a
virtual external field and leads to the condition
\begin{equation}
 \label{buk3400}
 \frac{\delta \ln Z_{M}}{\delta\phi_k(q_{n_m})} =0.
 \end{equation}

Namely, the condition of the extremum of the part of the functional written
in (\ref{buk31}) corresponds simultaneously to the condition that the sum in
brackets in (\ref{buk31}) tends to zero. Let us consider the proof of the
foregoing. We vary (\ref{buk31}) over all the potentials in (\ref{buk5})
assuming that they can all be independent of each other, which makes it
possible to substantially simplify the procedure of transformations.
Performing the variation with respect to an individual $\phi _{k}
\left(q_{n_{m} } \right)$ and using (\ref{buk7}) and (\ref{buk16}),
for the reference system part we obtain
\begin{align} \label{buk35}
\delta\ln z_{m}^{0} &=\frac {1}{z_{m}^{0}}\Bigg\{\int\limits_{\omega_{m}}\delta\phi_{k}\big(q_{0_{m}}\big)\exp\bigg[ -\beta\sum_{j(m)}^{Z}\phi_{j}
\big(q_{0_{m}}\big)\bigg] \rd q_{0_{m}} \nonumber \\
&+\int\limits_{\omega_{m}}\delta\phi\big(q_{1_{m}}\big)\exp\bigg[ -\beta\sum_{j(m)}^{Z}\phi_{j}
 \big(q_{1_{m}}\big)\bigg] {\rm e}^{\beta\mu_{m}} \rd q_{1_{m}}\Bigg\}.
 \end{align}

Since the variation in $\delta\ln Z_{M}$ is satisfied for fixed $\mu_{i}$, we
transform the second term in $Z_M$ into a form that contains explicit
$\mu_{i}$.
\begin{align} \label{buk36}
 \mathcal{I}_{m}=\frac{1}{2}\sum \limits _{j\left(m\right)}^{Z}
	\sum \limits _{n_{m} =0}^{1}\sum \limits _{n_{j} =0}^{1}\rho_{n_{m}} \rho_{n_{j} }
f_{n_{m} n_{j}}   =\frac{1}{2}\sum \limits _{j\left(m\right)}^{Z}\sum \limits
	_{n_{m} =0}^{1}\sum \limits _{n_{j} =0}^{1}\int \limits _{\omega _{m} }\int \limits
	_{\omega _{j} }\exp  \bigg(-\beta \Big\{\Phi \big(q_{n_{m} } ,q_{n_{j} } \big) \nonumber
\end{align}
\begin{align}
	+
	\sum \limits _{s\ne m,j}\left[\phi _{s} \big(q_{n_{m} } \big)+\phi _{s} \big(q_{n_{j}
	} \big)\right] \Big\}\bigg){ \text {e}^{\beta \mu _{m} n_{m} } \text {e}^{\beta
		\mu _{j} n_{j} }  \rd q_{n_{m} } \rd q_{n_{j} } \frac{1}{z_{m}^{0} z_{j}^{0} } } - \frac{1}{2}\,.  
\end{align}

Let us write the expression for the variation of (\ref{buk36}) with respect
to $\phi _{k} \left(q_{n_{m} } \right)$
\begin{align} \label{buk37} 
\delta\mathcal{I} _{m,k} &=\sum \limits _{j\ne
k,m}\sum \limits _{n_{m} =0}^{1}\sum \limits _{n_{j} =0}^{1}\int \limits _{\omega
_{m} }\int \limits _{\omega _{j} }\delta \phi _{k}      \left(q_{n_{m} } \right)
\exp \left[-\beta \Phi \big(q_{n_{m} } ,q_{n_{j} } \big)\right] 
 \nonumber \\
&\times \exp \Big\{-\beta \sum \limits _{s\ne m,j}\left[\phi _{s} \big(q_{_{m} } \big)+
\phi _{s} \big(q_{n_{j} } \big)\right] \Big\} \text {e}^{\beta \mu _{m} n_{m} } \text {e}^{\beta
\mu _{j} n_{j} } \rd q_{n_{m} } \rd q_{n_{j} } \displaystyle  \frac{1}{z_{m}^{0} z_{j}^{0} } 
\nonumber \\
&-\sum \limits _{j\left(m\right)}^{Z}\sum \limits _{n_{m} =0}^{1}\sum \limits
_{n_{j} =0}^{1}\int \limits _{\omega _{m} }\int \limits _{\omega _{j} }\exp      \Big\{-
\beta \Phi \big(q_{n_{m} } ,q_{n_{j} } \big)+\sum \limits _{s\ne m,j}\left[\phi
_{s} \left(q_{n_{m} } \right)+\phi _{s} \big(q_{n_{j} } \big)\right] \Big\}
 \nonumber \\
&\times  \text {e}^{\beta \mu _{m} n_{m} } \text {e}^{\beta \mu _{j} n_{j} }  \rd q_{n_{m} }
\rd q_{n_{j} }\displaystyle \frac{1}{z_{m}^{0} z_{_{j} }^{0} } \delta\ln z_{m}^{0} \,. 
\end{align}

The first sum on the right-hand side of equation (\ref{buk37}) does not
contain a term with $j= m$, which is convenient to add and subtract, which
allows the sum (\ref{buk35}) and (\ref{buk37}) to be represented in a form
that allows the separation of variables by cell numbers. Namely,
\begin{align} \label{buk38} 
\delta \ln z_{m}^{0} +\delta\mathcal{I} _{m,k}
&=\frac{1}{z^0_{m} } \sum \limits _{n_{m} =0}^{1}\int \limits _{\omega _{m} }
\delta \phi _{k}   \left(q_{n_{m} } \right)\exp \bigg[-\beta \sum \limits _{j\left(m
\right)}^{Z}\phi _{j} \left(q_{n_{m} } \right) \bigg] \text {e}^{\beta \mu _{m} n_{m} } \rd q_{n_{m}}
\nonumber \\
&-\frac{1}{z^0_{m} z^0_{k} } \sum \limits _{n_{m} =0}^{1}\sum \limits _{n_{k}
=0}^{1}\int \limits _{\omega _{k} }\delta \phi _{k}    \left(q_{n_{m} } \right) \rd q_{n_{m}
} \int \limits _{\omega _{k} } \rd q_{n_{k} }  \exp \left[-\beta \Phi \left(q_{n_{m}
} ,q_{n_{k} } \right)\right] 
\nonumber \\
&\times \exp \Big\{-\beta \sum \limits _{s
\ne m,k}\left[\phi _{s} \left(q_{n_{m} } \right)+\phi _{s} \left(q_{n_{k} } \right)
\right] \Big\}\exp \left[\beta \left(\mu _{m} n_{m} +\mu _{k} n_{k} \right)\right]
\nonumber \\
&+\sum \limits _{j\ne m}A_{m,j}  =0. 
\end{align}

In the relation (\ref{buk38}), the term $A_{m,j}$  is the symmetrized part of
$\delta\mathcal{I} _{m,k}$ determined from equation (\ref{buk37}). So, for
$A_{m,j}$ we can write
\begin{align} \label{buk39} 
A_{m,j} &=\displaystyle\frac{1}{z^0_{m} z^0_{j} } \sum
\limits _{n_{m} =0}^{1}\sum \limits _{n_{j} =0}^{1} \text {e}^{\beta \left(\mu _{m} n_{m}
+\mu _{j} n_{j} \right)}   \int \limits _{\omega _{m} }\int \limits _{\omega _{j}
}\left[\delta \phi _{k} \left(q_{n_{m} } \right)-\delta \ln z^0_{m} \right] 
\nonumber \\
&\times \exp \left[-\beta \Phi \big(q_{n_{m} } ,q_{n_{j} } \big)\right]\exp
\bigg\{-\beta \sum \limits _{s\ne m,j}\left[\phi _{s} \left(q_{n_{m} } \right)+\phi
_{s} \big(q_{n_{j} } \big)\right] \bigg\} \rd q_{n_{m} } \rd q_{n_{j} } \,. 
\end{align}

Subsequent separation of variables makes it possible to obtain an equation
for the required potentials $\phi _{j} \left(q_{n_{i} } \right)$. Since both
$\delta \phi _{k} \left(q_{1_{m} } \right)$ and $\delta \phi _{k}
\left(q_{0_{m} } \right)$ are independent, after simplifications of
equations (\ref{buk39}), we obtain a system of defining
equations
\begin{eqnarray} \label{buk40} 
\exp \left[-\beta \phi _{k} \left(q_{n_{m}
} \right)\right]=\frac{1}{z_{k}^{0} } \sum \limits _{n_{k} =0}^{1}
\text {e}^{\beta \mu _{k} n_{k}} \int \limits _{\omega _{k} } \rd q_{n_{k} }     
\exp \Bigg\{-\beta \bigg[\Phi \left(q_{n_{m} } ,q_{n_{k} } \right)+\sum \limits _{s
\ne m,k}^{Z}\phi _{s} \left(q_{n_{k} } \right) \bigg]\Bigg\} .
\end{eqnarray}

Equations (\ref{buk40}) can be rewritten in another form, namely when the
density field is used as a variable that defines the system. Substituting
(\ref{buk13}) into (\ref{buk40}) we arrive at a description
\begin{equation} \label{buk41} 
\exp \left[-\beta \phi _{k} \left(q_{n_{m}
} \right)\right]=\sum \limits _{n_{k} =0}^{1}\displaystyle\frac{\rho _{n_{k} } }{Q_{n_{k} } }
\int \limits _{\omega _{k} } \rd q_{n_{k} }  \exp \Bigg\{-\beta \bigg[
\Phi \left(q_{n_{m} } ,q_{n_{k} } \right)+\sum \limits _{s\ne m,k}^{Z}\phi _{s} \big(q_{n_{k}
} \big) \bigg]\Bigg\} . 
\end{equation}

Equations (\ref{buk40}) and (\ref{buk41}) define the single-particle cell
potentials under the condition that the two-vertex diagrams in the cluster
expansion of the original partition function (\ref{buk30}) are equal to zero.
As was shown in \cite{Argyrakis}, this condition leads to the results that are
equivalent to the quasichemical (or Bether-Peierls) approximation. The
inclusion of three-vertex diagrams in the equations (\ref{buk40}) or
(\ref{buk41}) is also possible but it would lead to more complicated
equations and, therefore, we will not consider it here.

Now, we show that when equation (\ref{buk40}) is satisfied, each term ${A_j}$
in (\ref{buk38}) turns out to be zero, which indicates that the separation
constant of the variables in (\ref{buk38}) chosen to be zero is correct.

Thus, substituting (\ref{buk40}) into (\ref{buk39}), we obtain
\begin{equation} \label{buk42}
 A_{m,j} =\frac{1}{z^0_{m} } \sum \limits _{n_{m=0} }^{1} \text {e}^{
\beta \mu _{m} n_{m} }  \int \limits _{\omega _{m} }\left[\delta \phi _{k} \left(q_{n_{m}
} \right)-\delta \ln z^0_{m} \right]  \exp \Big[-\beta \sum \limits _{s\ne m}
\phi _{s} \left(q_{n_{m} } \right) \Big] \rd q_{n_{m} }\,.  
\end{equation}

Using the definitions (\ref{buk9}) and (\ref{buk10}) in (\ref{buk42}),
we obtain
\begin{equation} \label{buk43}
A_{m,j} =\frac{1}{z^0_{m} } \sum \limits
_{n_{m=0} }^{1} \text {e}^{\beta \mu _{m} n_{m} }  \int \limits _{\omega _{m} }\delta \phi
_{k}  \left(q_{n_{m} } \right)  \exp \Big[-\beta \sum \limits _{s
\ne m}\phi _{s} \left(q_{n_{m} } \right) \Big] \rd q_{n_{m} } -\delta \ln z_{m}^{0} \,.
\end{equation}

In turn, it is clear from (\ref{buk35}) that the first sum in (\ref{buk43})
is identical to the variation of the second term, which proves that $A_j=0$
when choosing single-particle potentials satisfying equation (\ref{buk40}).
As already noted for specific calculations, (\ref{buk40}) is preferable in
the form (\ref{buk41}) because this option allows for an implicit
relationship between the chemical potential and the density to be replaced by
an explicit one. Namely, for the chosen $\rho _{n_{i} } $ the solution
(\ref{buk41}) is found with (\ref{buk10}) taken into account. Then, due to~(\ref{buk15}), the chemical potential and the free energy of the system are
determined. To calculate the latter, it is convenient to use the relation
\begin{equation} \label{buk44} 
-\beta F=\sum \limits _{i=1}^{M}\big(\ln z^0_i-\rho _{1_{i}
} \mu _{i} \big).  
\end{equation}

Taking into account equation (\ref{buk15}), expression (\ref{buk44}) is
represented in the form
\begin{equation} \label{buk45} 
-\beta F=\sum \limits _{i=1}^{M}\left[\ln z^0_{i} -\rho
_{1_{i} } \ln \left(\frac{\rho _{1_{i} } Q_{0_{i} } }{\rho _{0_{i} } Q_{1_{i} } }
\right)\right].  
\end{equation}

As a result, it follows from equation (\ref{buk40}) that
\begin{equation} \label{buk46} 
\sum \limits _{n_{j} =0}^{1}\rho _{n_{j} }  f_{n_{i}
n_{j} } =0. 
\end{equation}

In order to prove (\ref{buk46}), we multiply equation (\ref{buk41}) by
\begin{equation} \label{buk45a}
\frac{\rho _{n_{m} } }{Q_{n_{m} } } \exp \Big[-\beta \sum \limits _{s\ne m}\phi
_{s} \left(q_{n_{m} } \right) \Big].
\end{equation}

After integration, we obtain
\begin{equation} \label{buk45ab}
1=\sum \limits _{n_{m} =0}^{1}\sum \limits _{n_{k} =0}^{1}\rho
_{n_{m} }   \rho _{n_{k} } \int \limits _{\omega _{m} }\int \limits _{\omega _{k}
}\exp \left[-\beta \Delta \phi \left(q_{n_{m} } ,q_{n_{k} } \right)\right]
 \widehat{F}_{11} \left(q_{n_{m} } \right)\widehat{F}_{11} \left(q_{n_{k}
} \right) \rd q_{n_{m} } \rd q_{n_{k} } \,.
\end{equation}

The identity obtained with the definition of (\ref{buk32}) proves the
validity of equation (\ref{buk46}). Hence, it follows that when the potentials
are determined from (\ref{buk41}), the results of all three approaches are
the same. Due to (\ref{buk46}), $\ln z_{i} $ coincides with $\ln z_{i}^{0}$
which is defined by the relation (\ref{buk16}).

 Substituting (\ref{buk16}) into the formula (\ref{buk45}), we obtain an
expression for the free energy of the system in the form
\[-\beta F=\sum \limits _{i=1}^{M}\left(\ln Q_{0_{i} } -\ln \rho _{0_{i} } +\rho
_{1_{i} } \ln \rho _{0_{i} } +\rho _{1_{i} } \ln Q_{1_{i} } -\rho _{1_{i} } \ln \rho
_{1_{i} } -\rho _{1_{i} } \ln Q_{0_{i} } \right) \]
or
\begin{equation} \label{buk47} -\beta F=\sum \limits _{i=1}^{M}\left(\rho _{0_{i} }
\ln Q_{0_{i} } +\rho _{1_{i} } \ln Q_{1_{i} } -\rho _{0_{i} } \ln \rho _{0_{i} }
-\rho _{1_{i} } \ln \rho _{1_{i} } \right).  \end{equation}

Such a representation of the right-hand side of equation (\ref{buk47})
corresponds to a configurational integral in the form
\begin{equation} \label{buk48} Q_{N} =Q_{N}^{0} =\sum \limits _{i=1}^{M}\frac{Q_{1_{i}
}^{\rho _{1_{i} } } Q_{0_{i} }^{\rho _{0_{i} } } }{\rho _{1_{i} }^{\rho _{1_{i} }
} \rho _{0_{i} }^{\rho _{0_{i} } } }\,. \end{equation}

\section{Verification of thermodynamic self-consistency of the theory}\label{sec5}

The expression for $Q_{N}^{0} $ in the form (\ref{buk48}) can be obtained
using the Hamiltonian
\begin{equation} \label{buk49} H_{N}^{0} =\sum \limits _{i=1}^{M}\sum \limits _{j\left(i
\right)}^{Z}\phi _{j}   \big(q_{n_{i} } \big). \end{equation}

To this end, it is necessary to consider the states corresponding to
(\ref{buk49}) based on the methods for forming local equilibrium
distributions \cite{Zubarev71}. We show that the conditions (\ref{buk41}),
when the contributions of the third and the subsequent virial coefficients
are not taken into account, give the identical $\rho_i$ determined by
formulae
\begin{equation} \label{buk50} \mu _{i} =\frac{\partial F}{\partial \rho _{i} }  \end{equation}
and
\begin{equation}
\label{buk51} \rho _{1_{i} } =\frac{\partial \ln Z_{M}^{0} }{\partial \left(\beta\mu _{i}
\right)}.  \end{equation}

In order to verify the thermodynamic consistency of the theory, we first
consider the validity of equation (\ref{buk51}). Using (\ref{buk16}), we
obtain
%
\begin{align} \label{buk52} 
\rho _{1_{i} } &=\frac{\partial \ln
		\left(Q_{0_{i} } + \text {e}^{\beta \mu _{i} } Q_{1_{i} } \right)}{\partial \left(\beta \mu
		_{i} \right)} +\sum \limits _{k\ne i}^{M}\frac{\partial \ln z_{k}^{0} }{\partial
		\left(\beta \mu _{i} \right)}
\nonumber \\
&=\frac{1}{\left(Q_{0_{i} } +\text {e}^{\beta \mu _{i}
		} Q_{1_{i} } \right)} \left[\text {e}^{\beta \mu _{i} } Q_{1_{i} } +\frac{\partial Q_{0_{i} } }{
		\partial \left(\beta \mu _{i} \right)} +\frac{\partial Q_{1_{i} } }{\partial \left(
		\beta \mu _{i} \right)} \right]  +\sum \limits _{k\ne 1}^{M}\frac{\partial \ln
		z_{k}^{0} }{\partial \left(\beta \mu _{i} \right) }
\nonumber \\
&=\rho _{1_{i} } +\frac{1}{\left(Q_{0_{i} } +\text {e}^{
			\beta \mu _{i} } Q_{1_{i} } \right)} \left[\frac{\partial }{\partial \left(\beta
		\mu _{i} \right)} Q_{0_{i} } +\text {e}^{\beta \mu _{i} } \frac{\partial Q_{1_{i} } }{\partial
		\left(\beta \mu _{i} \right)} \right]+ \sum \limits _{k\ne i}^{M}\frac{\partial
		\ln z_{k}^{0} }{\partial \left(\beta \mu _{i} \right) }\,. 
\end{align}

It follows from (\ref{buk52}) that (\ref{buk40}) must satisfy the additional
condition
\begin{equation} \label{buk53} \frac{\partial }{\partial \left(\beta \mu _{i} \right)}
Q_{0_{i} } +\text {e}^{\beta \mu _{i} } \frac{\partial Q_{1_{i} } }{\partial \left(\beta
\mu _{i} \right)} =0. \end{equation}

Then, $\rho_{1_{i} }$ calculated from the formulae (\ref{buk51}) and
(\ref{buk14}) will coincide. To prove (\ref{buk53}) we differentiate the
condition (\ref{buk40}) with respect to $\beta{\mu_{m}}$
\begin{eqnarray} \label{buk54} 
&&\frac{\partial }{\partial \left(\beta
\mu _{m} \right)} \left[-\beta \phi _{k} \left(q_{n_{m}} \right)\right]\exp \left[-
\beta \phi _{k} \left(q_{n_{m} } \right)\right]
=-\frac{1}{z_{k}^{0} } \frac{
\partial \ln z_{k}^{0} }{\partial \left(\beta \mu _{m} \right)} \sum \limits _{n_{k}
=0}^{1}\exp \left(\beta \mu _{k} n_{k} \right)
\nonumber \\
&&\times \int \limits _{\omega _{k} } \rd q_{n_{k}
}   \exp \bigg\{-\beta \bigg[\Phi \left(q_{n_{m} } ,q_{n_{k} } \right)+
\sum \limits _{s\ne m,k}^{Z}\phi _{s} \left(q_{n_{k} } \right) \bigg]\bigg\} 
+\frac{1}{z_{k}^{0} } \sum \limits _{n_{k} =0}^{1}\exp \left(\beta \mu _{k} n_{k}
\right)
\nonumber \\
&&\times \int \limits _{\omega _{k} }\left[\sum \limits _{s=m,k}^{Z}\frac{\partial
\phi _{s} \left(q_{n_{k} } \right)}{\partial \left(\beta \mu _{m} \right)}  \right]
\exp \bigg\{-\beta \bigg[\Phi \left(q_{n_{m} } ,q_{n_{k} } \right)+
\sum \limits _{s\ne m,k}^{Z}\phi _{s} \left(q_{n_{k} } \right) \bigg]\bigg\} \rd q_{n_{k}
}. 
\end{eqnarray}

Multiplying (\ref{buk43}) by $\displaystyle\frac{1}{z_{m}^{0} } \exp \bigg[-\beta \sum
\limits _{s\ne k,m}\phi _{s} \left(q_{n_{m} } \right) \bigg]$, after
integration over $q_{n_{m} } $ we obtain
\begin{eqnarray} \label{buk55} 
&&\frac{1}{z_{m}^{0} } \int \limits
_{\omega _{m} }\frac{\partial }{\partial \left(\beta \mu _{m} \right)}  \left[-\beta
\phi _{k} \left(q_{n_{m} } \right)\right]\exp \left[-\beta \sum \limits _{s\ne k}^{Z}
\phi _{s} \left(q_{n_{m} } \right) \right] \rd q_{n_{m} } 
\nonumber \\
&&=-\frac{1}{z_{m}^{0}
} \frac{\partial \ln z_{k}^{0} }{\partial \left(\beta \mu _{m} \right)} \int \limits
_{\omega _{m} }\exp \left[-\beta \sum \limits _{s\ne k}^{Z}\phi _{s} \left(q_{n_{m}
} \right) \right] \rd q_{n_{m} } 
\nonumber \\
&&+\frac{1}{z_{k}^{0} z_{m}^{0} } \sum \limits
_{n_{k} =0}^{1}\exp \left(\beta \mu _{k} n_{k} \right) \int \limits _{\omega _{k}
}\int \limits _{\omega _{m} } \rd q_{n_{m} }   \rd q_{n_{k} } \exp \left[-\beta \Phi \left(q_{n_{k}
} ,q_{n_{m} } \right)\right] 
\nonumber \\
&&\times \left[-\beta \sum \limits _{s\ne m,k}^{Z}
\displaystyle\frac{\partial \phi _{s} \left(q_{n_{k} } \right)}{\partial \left(\beta \mu _{m}
\right)}  \right]\exp \left[-\beta \sum \limits _{s\ne k}^{Z}\phi _{s} \left(q_{n_{m}
} \right) \right]. 
\end{eqnarray}

Multiplying (\ref{buk55}) by $\sum \limits _{n_{m} }^{1} \text {e}^{-\beta \mu _{m}
n_{m} }$ and taking into account (\ref{buk9}) and (\ref{buk40}), we have
\begin{eqnarray} \label{buk56} 
&&\frac{2}{z_{m}^{0} } \sum \limits
_{n_{m} =0}^{1}\text {e}^{-\beta \mu _{m} n_{m} }  \int \limits _{\omega _{m} }\frac{\partial
}{\partial \left(\beta \mu _{n} \right)}  \left[-\beta \phi _{k} \left(q_{n_{m} }
\right)\right]\exp \left[-\beta \sum \limits _{s\ne k}^{Z}\phi _{s} \left(q_{n_{m}
} \right) \right] \rd q_{m}  
\nonumber \\
&&=-\frac{\partial \ln z_{k}^{0} }{\partial \left(\beta
\mu _{m} \right)} +\frac{1}{z_{k}^{0} } \sum \limits _{n_{k} =0}^{1}\exp  \left(
\beta \mu _{k} n_{k} \right)\int \limits _{\omega _{k} }\left[-\beta \sum \limits
_{S\ne k}^{Z}\frac{\partial \phi _{s} \left(q_{n_{k} } \right)}{\partial \left(\beta
\mu _{m} \right)}  \right] \exp \left[-\beta \sum \limits _{s\ne
k}^{Z}\phi _{s} \left(q_{n_{m} } \right) \right] \rd q_{n_{k} }. 
\end{eqnarray}

As a result,
\begin{eqnarray}
\label{buk57} 
&&\frac{2}{z_{m}^{0} } \sum \limits _{n_{m} =0}^{1}\text {e}^{-
\beta \mu _{m} n_{m} }  \int \limits _{\omega _{m} }\frac{\partial }{\partial \left(
\beta \mu _{m} \right)}  \left[-\beta \phi _{k} \left(q_{n_{m} } \right)\right]
 \exp \left[-\beta \sum \limits _{s\ne k}^{Z}\phi _{s} \left(q_{n_{m}
} \right) \right] \rd q_{n_{m} }
\nonumber \\
&&=\frac{\partial \ln z_{k}^{0} }{\partial \left(
\beta \mu _{m} \right)} +\frac{1}{z_{k}^{0} } \sum \limits _{n_{k} =0}^{1}\exp \left(
\beta \mu _{k} n_{k} \right) \frac{\partial Q_{n_{k} } }\partial \left(\beta \mu
_{m} \right)=0. 
\end{eqnarray}

Performing the summation in (\ref{buk57}) with respect to $k\neq m$, we find
\begin{equation}
\label{buk58} \sum \limits _{n_{m} =0}^{1}\text {e}^{-\beta \mu _{m} n_{m} }  \frac{\partial
Q_{n_{m} } }{\partial \left(\beta \mu _{m} \right)} =0. \end{equation}

Likewise, we show that although in accordance with (\ref{buk40}) each $\phi
_{s} \left(q_{n_{k} } \right)$ depends on the chemical potentials in the
entire range of the $k$-th node, it turns out that
\begin{equation} \label{buk59} \frac{\partial z^0_{m} }{\partial \left(\beta \mu _{l}
\right)} =0 \end{equation}
for $l\neq m$.

In combination with the condition (\ref{buk58}), this will ultimately prove
the thermodynamic consistency of the theory when calculating the
thermodynamic potentials $F$ and $\Omega$ and their first derivatives,
determined by formulae (\ref{buk15}) and (\ref{buk51}). The proof of the
condition (\ref{buk59}) is carried out similarly to (\ref{buk53}).

In order to reduce the transformations and make them more transparent, we
write a relation analogous to (\ref{buk54}) in the form
\begin{eqnarray} \label{buk60} 
&&\bigg\{\displaystyle\frac{\partial }{\partial \left(
\beta \mu _{s} \right)} \left[-2\beta \phi _{k} \left(q_{n_{m} } \right)\right]\bigg\}
\exp \left[-\beta \phi _{k} \left(q_{n_{m} } \right)\right]=\displaystyle-\frac{\partial
\ln z_{k}^{0} }{\partial \left(\beta \mu _{s} \right)} \exp \left[-\beta \phi _{k}
\left(q_{n_{m} } \right)\right] 
\nonumber \\
&&\displaystyle+\frac{1}{z_{k}^{0} } \sum \limits _{n_{k}
=0}^{1}\exp  \left(\beta \mu _{k} n_{k} \right)    \int \limits _{
\omega _{k} }\left[\sum \limits _{s\ne m}^{Z}\displaystyle\frac{\partial \beta \phi _{s} \left(q_{n_{k}
} \right)}{\partial \left(\beta \mu _{s} \right)}   \right]\exp \Bigg\{-\beta
\bigg[\Phi \left(q_{n_{m} } ,q_{n_{k} } \right)
\nonumber \\
&&+\sum \limits _{s\ne m,k}^{Z}\phi
_{s} \left(q_{n_{k} } \right) \bigg]\Bigg\}  \rd q_{n_{k} } . 
\end{eqnarray}

Let us apply the following operation to
relation (\ref{buk60})
\begin{equation} \label{buk61} \widehat{\pi }=\sum \limits _{k\ne m}^{Z}\frac{1}{z_{m}^{0}
}  \sum \limits _{n_{m} }^{1}\exp \left(-\beta \mu _{m} n_{m} \right) \int \limits
_{\omega _{m} }\exp \left[-\beta \sum \limits _{s\ne k}^{Z}\phi _{s} \left(q_{n_{m}
} \right) \right] \rd q_{n_{m} }  ... \end{equation}
writing equation (\ref{buk60}) in a symbolic form
\begin{equation}
\label{buk62}
A=B+C.
\end{equation}

When converting the left-hand side of expression (\ref{buk62}), the summation
sign with respect to $k \neq m$ is put under the integration sign with respect to $\omega_m $,
and taking into account (\ref{buk10}), we write the result in the form
\begin{equation} \label{buk63} \widehat{\pi }A=2\frac{1}{z_{m}^{0} } \sum \limits _{n_{m}
=0}^{1}\exp \left(-\beta \mu _{m} n_{m} \right) \frac{\partial Q_{n_{m} } }{\partial
\left(\beta \mu _{s} \right)}.  \end{equation}

Since in (\ref{buk63}) $s \neq m$ and $\mu_m $ and  $\mu_s $ are independent
variables, we switch the summation over $n_{m}$ and differentiation with
respect to $\mu_s$, which due to (\ref{buk9}) allows us to write
\begin{equation} \label{buk64} \widehat{\pi }A=2\frac{\partial \ln z_{m}^{0} }{\partial
\left(\beta \mu _{s} \right)}.  \end{equation}

The result of the transformation of the expression $B$ is equal to 1. It can be
obtained if the operation~(\ref{buk51}) is applied to $\exp \left[-\beta \phi
_{k} \left(q_{n_{m} } \right)\right]$ taking the constant as the sign of
operation (\ref{buk61}). As a result, direct application of $\widehat{\pi }$
gives
\begin{equation} \label{buk65} \widehat{\pi }B=-\frac{\partial \ln z_{k}^{0} }{\partial
\left(\beta \mu _{s} \right)}.  \end{equation}

To transform the expression $C$, we switch the summation and integration
operations in (\ref{buk61}), place the operator $\widehat{\pi }$ under the
integration sign with respect to $\omega_k$ and take into account that
according to (\ref{buk40})
\begin{equation} \label{buk66} \widehat{\pi }\exp \big\{-\beta \left[\Phi \left(q_{n_{k}
} ,q_{n_{m} } \right)\right]\big\}=\exp \left[-\beta \phi _{m} \left(q_{n_{k} }
\right)\right]. \end{equation}

Then, $\widehat{\pi }C$ becomes of the form
\begin{equation} \label{buk67} \widehat{\pi }C=\frac{1}{z_{k}^{0} } \sum \limits _{n_{k}
=0}^{1}\exp \left(\beta \mu _{k} n_{k} \right) \frac{\partial \ln Q_{k} }{\partial
\left(\beta \mu _{s} \right)}.  \end{equation}

Taking (\ref{buk9}) into account, condition (\ref{buk67}) gives
\begin{equation} \label{buk68} \widehat{\pi }C=\frac{\partial \ln z_{k}^{0} }{\partial
\left(\beta \mu _{s} \right)}.  \end{equation}

The results (\ref{buk65}) and (\ref{buk68}) show that according to
(\ref{buk62}), the expression (\ref{buk64}) is identically equal to zero.
This, in turn, proves the validity of (\ref{buk60}).

Similarly to the consistency of (\ref{buk51}) and (\ref{buk13}) proved by
derivations (\ref{buk52})--(\ref{buk68}), one can show that (\ref{buk50}) is
consistent with (\ref{buk14}). To this end, it is necessary to perform
transformations similar to those performed for (\ref{buk40}) over
(\ref{buk41}), differentiating each of the equations
(\ref{buk52})--(\ref{buk68}) with respect to thermodynamic variables $\rho_{n_{i}}$ but not as it was before with respect to $\mu_{s}$.
Further analysis of the consistency problem involves comparing the results of
calculating the second derivatives of thermodynamic potentials, and will be
considered separately.

\section{Discussion}\label{sec6}

In this paper, for a lattice model with the Hamiltonian (\ref{buk23}) and a
pair interaction potential $\Phi(q_{n_i},q_{n_j})$ we formulate a reference
system with the Hamiltonian (\ref{buk1}) and a single-particle cell potential
$\phi_j(q_{n_i})$. The potential $\phi_j(q_{n_i})$ can be interpreted as the
mean potential exerted by a particle in the lattice cell $\omega_j$ on a
particle in the lattice cell $\omega_i$. It is shown that the system with the
single-cell potential $\phi_j(q_{n_i})$ reduces to the description of a
Fermi-like lattice model in an external field. Using this system as the
reference system and renormalized Mayer functions in the form (\ref{buk29})
with $\Delta \phi (q_{n_{i} } ,q_{n_{j} })$ in the form
(\ref{buk26}), the generalized cluster expansion for the free energy of the
considered system is obtained. The cell potentials are calculated from the
condition of the minimum difference of thermodynamic properties of the
systems with the Hamiltonians (\ref{buk23}) and (\ref{buk1}). Such a
procedure is considered under the condition that the two-vertex diagrams in
the cluster expansion of the partition function for the system with the
Hamiltonian~(\ref{buk23}) are equal to zero. As a result, for the
single-particle cell potential, a system of equations~(\ref{buk40}) was
obtained. This system can also be presented in the form of equation
(\ref{buk41}). As it was noted previously in the reference \cite{Argyrakis},
such a description is equivalent to the quasichemical approximation. We
should note that the descriptions of the considered system in the framework
of the Hamiltonians (\ref{buk23}) and (\ref{buk1}) are exactly equivalent
only in the case when all the terms in the cluster expansion (\ref{buk33})
are taken into account at the calculation of the single-particle potentials.
There are two principal differences between the considered equation
(\ref{buk40}) and the corresponding equations in traditional approaches such
as the mean field approximation formulated in the framework of the field
theoretical approach \cite{DiCaprio11,Kravtsiv15} or the density functional
approach \cite{Evans1992} well developed for non-lattice fluid systems. The
first difference is connected with the presence of the interparticle
potential $\Phi (q_{n_{i} } ,q_{n_{j} })$ in equation (\ref{buk40})
in the exponential form. The second one is connected with the inequality
$s\ne m,k$ in the exponent of equation (\ref{buk40}). It means that instead
of the singlet distribution function $\hat{F}_{11}(q_{n_k})$, which usually
appears in the mean field approximation, in the approach considered, the
function $\hat{F}_{11}(q_{n_k})\exp\left[\beta\phi_m(q_{n_k})\right]$
appears. This is the result of the peculiarity of the Mayer functions in the
form (\ref{buk29}) with $\Delta \phi (q_{n_{i} } ,q_{n_{j} } )$ in
the form (\ref{buk26}). If we neglect the condition $s\ne m,k$ due to the
renormalization condition (\ref{buk19}), the equation (\ref{buk40}) can be
rewritten in the form
\begin{equation}
\text {e}^{-\beta\phi_k(q_{n_m})}=1+\sum\limits_{n_k=0}^1\rho_{n_k}\int \rd q_{n_k}\big\{\exp\left[-\beta
\Phi\left(q_{n_{m} } ,q_{n_{k} }\right)\right]-1\big\}\hat{F}_{11}(q_{n_k})\,,
\end{equation}
which after linearization of the exponents
$\exp\left[{-\beta\phi_k(q_{n_m})}\right]$ and $\exp\left[-\beta
\Phi\left(q_{n_{m} } ,q_{n_{k} }\right)\right]$ leads to a traditional form
for the mean field approximation
\begin{equation}
\phi_k(q_{n_m})=\sum\limits_{n_k=0}^1\rho_{n_k}\int \rd q_{n_k}
\Phi\left(q_{n_{m} } ,q_{n_{k} }\right)\hat{F}_{11}\left(q_{n_k}\right).
\end{equation}

The equation (\ref{buk40}) describes the single-particle cell potentials
$\phi_k(q_{1_m})$ for real particles, but for vacancies it is more of a
problem. This problem is similar to the description of solvophobic
interaction in the theory of solutions \cite{Ronis1977,Bandura2018} and for a
correct description of single-particle cell potential $\phi_k(q_{0_m})$ for
vacancies, at least three-vertex diagrams in the equation (\ref{buk40}) should
be included. In the two-vertex diagram approximation for $\phi_k(q_{0_m})$
due to inequality $s\ne m,k$ there appears only some constant corresponding
to the change of the chemical potential due to the creation of a vacancy.

The theory presented here is easily generalized when it is necessary to take
into account a larger number of possible states. This is achieved by
expanding the possible values of the occupation numbers, when
$n_i=0,1,2\dots$.
In this case, for every microconfiguration given by the set of values
$\{n_1,\dots ,n_m\}$, the expression for the Hamiltonian is completely
preserved in the form (\ref{buk23}). So, for example, for $n = 0,1,2$ in all
formulae (\ref{buk1})--(\ref{buk68}) one should remember that
\begin{eqnarray} \label{buk69} 
\Phi \big(q_{n_{i} }, q_{n_{j}
} \big)=\left\{
\begin{array}{l} 
{0,\, n_{i},  n_{j} =0}\,, 
\\ 
{h\left(q_{i} ,q_{j}
\right),\, n_{i},  n_{j} =1}\,, 
\\
{h\big(q_{i} ,q_{j} \big)+h\big(q_{i} ,q'_{j}
\big),\, n_{i},n_{j} =2}\,, 
\\
{h\left(q_{i} ,q_{j} \right)+h\big(q_{i} ,q'_{j}
\big)+h\big(q'_{i} ,q_{j} \big)+h\big(q'_{i} ,q'_{j} \big),\, n_{i},
n_{j} =4}\,. 
\end{array} \right.
\end{eqnarray} 
Here, $h(q_{i} ,q_{j})$ is the intermolecular interaction
potential of two particles in positions $q_{i} $ and $q_{j} $, when
$(q_{i} ,q'_{i}) \in \omega _{i} $, $(q_{j} ,q'_{j}
)\in \omega _{j} $.

In addition to taking into account (\ref{buk69}) in all expressions
(\ref{buk1})--(\ref{buk68}), the summation over $n = 0,1$ must be extended
to the case $n = 0,1,2$, additionally taking into account that
\begin{equation} \label{buk70} \int \limits _{\omega _{i} } \rd q_{0_{i} }  =
\frac{1}{\omega _{i} } \int \limits _{\omega _{i} } \rd q_{i}  ,\quad \int \limits _{
\omega _{i} }\rd q_{1_{i} }  =\int \limits _{\omega _{i} } \rd q_{i}  ,\quad \int \limits
_{\omega _{i} } \rd q_{2_{i} }  =\int \limits _{\omega _{i} } \rd q_{i}  \int \limits _{
\omega _{i} } \rd q'_{i}\,.   \end{equation}

Thus, it is shown that the properties of a condensed system can be described
by combining the model of an ideal crystal with a group expansion over the
modified Mayer functions that impose correlations on the properties of an
ideal crystal.

 It seems justified to extend the developed approach to take into account long-range
effects in essentially inhomogeneous media. Let us demonstrate the
possibility of such propagation using the example when the inhomogeneity of
the medium is described by different sizes of microcells, provided that in
each of them there is one particle.
 Representing
the energy of the system by short-range $\Phi (i,j)$ and long-range $V(i,j)$
potentials of pair interactions, respectively, for particles in positions
$q_{i} $, $q_{j} $, we present the configurational integral $Q_{N} $ of the
system in the form
\begin{equation}
\label{buk71}
Q_{N} =Q_{N}^{0} \bigg\langle \exp \bigg[-\beta \sum \limits _{i<j}^{N}V(i,j) \bigg]\prod
\limits _{i<j}^{N}\big[1+f(i,j)\big] \bigg\rangle _{0}\,,
\end{equation}
where
\begin{equation}
\label{buk72}
f(i,j)=\exp \big\{-\beta \left[\Phi \left(i,j\right)\right]-\phi _{j} \left(i\right)-
\phi _{i} \left(j\right)\big\}-1\,,
\end{equation}
\begin{equation}
\label{buk73}
 Q_{N}^{0} =\prod \limits _{i=1}^{N}Q_{i}\,,  {\rm \; \; \; \; \; \; \; \; }Q_{i}
=\int \limits _{v_{i} }\exp \Big[-\beta \sum \limits _{k\ne i}\phi _{k} \left(i
\right) \Big] \rd q_{i}\,.
\end{equation}
$Q_{N}^{0} $ is the configurational integral of an ideal crystal, expressed
through single-particle cell potentials of mean forces $\phi _{j}
\left(i\right)$, $f(i,j)$ is a renormalized Mayer function, the angle
brackets $\left\langle ...\right\rangle _{0} $ denote averaging over the
equilibrium states of the reference system.
The subsequent cumulant expansion of the expression~(\ref{buk1}) with respect
to the functions (\ref{buk2}) allows one to write in the approximation of the
second virial coefficient

\begin{equation}
\label{buk74d}\ln Q_{N} =\ln Q_{L}^{0} +\ln Q_{N}^{0} +\sum \limits _{i,j}\left\langle f(i,j)g(i,j)
\right\rangle _{0}  +...\,,
\end{equation}

\begin{equation}
\label{buk74}
 Q_{L}^{0} =\Big\langle \exp \Big[-\beta \sum \limits _{i<j}V(i,j) \Big]\Big
\rangle _{0} .
\end{equation}

The averaging in relations (\ref{buk73}) and (\ref{buk74}) is realized by
multiplying the unary distribution functions $F_{11} (i)$ and $F_{11} (j)$
\begin{equation}
\label{buk75}
 F_{11}
\left(i\right)=\frac{1}{Q_{i} } \exp \Big[-\beta \sum \limits _{k\ne 1}\phi _{k}
(i) \Big].
\end{equation}

A special feature of relation (\ref{buk74}) is that the renormalized Mayer
function $f(i,j)$ is modulated here by a binary function, for which a
consistent calculation scheme was developed in \cite{Yukhnovsky80}.
Applying the procedure of self-consistent calculation of the potentials $\phi _{j}
\left(i\right)$ in accordance with the foregoing, we arrive at a closed system of
equations of the form
\begin{equation} \label{buk76}\exp \left[-\beta \phi _{j} \left(i\right)\right]=\frac{1}{Q_{j} } \int \limits
_{v_{j} }g(i,j)\exp  \left[-\beta \Phi \left(i,j\right)\right]
 \times \exp \Big[-\beta \sum \limits _{k\ne
i,j}\phi _{k} \left(j\right) \Big] \rd j. \end{equation}

The system of equations (\ref{buk76}) differs from the one used earlier because its kernel, in addition to the point short-range potential, contains a binary
function for a system of particles with a Coulomb interaction, the expression for
which has the form \cite{Yukhnovsky80}
\begin{align}
F_{2} (i,j)&=F_{11} (i)F_{11} (j)g(i,j)\,,\nonumber\\
g(i,j)&=\exp [-\beta u(i,j)]\,,\label{buk77}
\end{align}
where $u(i,j)$ is a screening potential. Such a form solves the problem of
the divergence of the integrals when calculating the free energy
(\ref{buk73}). As a result, long-range effects apply also to the
re\-nor\-ma\-li\-za\-tion of single-particle cell potentials. In expression
(\ref{buk77}), only the initial terms of the series are written corresponding
to the Debye description of ion systems. A more complete representation for
relation~(\ref{buk77}) follows when calculating additional terms of the
distribution function $g(i,j)$ using collective variables~\cite{Yukhnovsky80}. A generalization of Debye screening to the system of
mobile ions in lattice models was recently discussed in \cite{Bokun18}. It
leads to the change of the traditional inverse Debye length $\kappa$ to the
new one $\kappa=\left[\beta
e^2c\left(1-c\right)/\left(\varepsilon_0\varepsilon h^3\right)\right]^{1/2}$,
where $e$ is the charge of a mobile ion, $\varepsilon_0$ is the dielectric
permittivity of the vacuum, $\varepsilon$ is the relative dielectric
permittivity of the medium, $c$ is the concentration of mobile ions, $h$ is
the length of the cell.

The peculiarity of the equations in the case under consideration is due to
the fact that the function $g(i,j)$ is determined by averaging not over the
states of an ideal gas, but over the states of an ideal crystal. Accordingly,
for example for ${i} = 1$, ${j} = 2$, the function $g(i,j)$ is defined by an
expression of the form
 \begin{eqnarray} \label{buk78}
g(1,2)=\int \limits _{v_{3} }...\int \limits _{v_{N} }\exp \bigg[-\beta \sum \limits
_{l<m=1}^{N}V\left(l,m\right) \bigg] 
 F_{11} \left(3\right)F_{11} \left(4\right)...F_{11} \left(N\right) \rd 3 \rd 4...\rd N.
 \end{eqnarray}

There are reasons to believe that the appearance of additional Gaussian in
(\ref{buk78}) introduced by the functions $F_{11}$ will lead to an
improvement in the convergence of the series (\ref{buk77}) obtained in the
framework of the collective variables method \cite{Yukhnovsky80} since a
similar procedure was successful in constructing the description using an
effective potential \cite{Ma76}.

\section{Conclusions}\label{sec7}

A method for modifying the cluster virial expansion, which makes it possible
to describe condensed state, is proposed. The method is based on replacing
the averaging over the states of an ideal gas by the averaging over the
states of an ideal crystal. The approach outlined differs from the well-known
perturbation theory because the basis with respect to which the expansion is
performed is not a part of the original Hamiltonian but is introduced
independently. The basis distribution conveys the main features of the solid
state of matter, where the motion of molecules is of an oscillatory nature
with respect to the lattice sites. As a result, the Hamiltonian of the basic
reference system is represented by the sum of single-particle cell mean force
potentials. It is shown how the Mayer functions can be modified so that they
would act as a small parameter for the subsequent expansion of the
thermodynamic potential by cumulant expansions. From the condition of
independence of the initial partition function on the introduced potentials
of the mean forces, a system of integral equations determining the above
potentials is obtained. The expression for the free energy functional is
obtained with the two first terms of its expansion in correlations taken into
account. It is shown that the obtained equations satisfy several optimization
conditions for the parameters characterizing the properties of the reference
system and the initial system. The thermodynamic consistency of various
methods for calculating the thermodynamic characteristics of a condensed
medium both in the canonical and in the grand canonical ensembles is proved.
A possibility of using the developed approach to take into account not only
short-range but also long-range interactions is shown.

\section*{Acknowledgements}

This project has received funding from European Unions Horizon 2020 research
and innovation programme under the Marie Sk\l{}odowska-Curie (grant agreement No
734276), the Belarusian Republican Foundation for Fundamental Research (grant
No $\Phi$16K-061) and the State Fund for Fundamental Research of Ukraine
(grant No $\Phi$73/26-2017). We thank Ivan Kravtsiv and Dung di Caprio for
the careful reading of the manuscript and useful comments.

\ukrainianpart

\title{Групове розвинення для опису конденсованих систем: підхід кристалічних комірок}
\author{Г.С. Бокун\refaddr{label1}, М.Ф. Головко\refaddr{label2}}
\addresses{
	\addr{label1} Білоруський державний технологічний університет, вул. Свердлова, 13а, 220006 Мінськ, Білорусь
	\addr{label2} Інститут  фізики конденсованих систем НАН України, вул. Свєнціцького, 1, 79011 Львів, Україна
}
%
%
%

\makeukrtitle

\begin{abstract}
	\tolerance=3000%
	Широко відоме групове розвинення, яке приводить до віріального розвинення
для вільної енергії розріджених систем, модифіковано так, щоб його можна було
застосовувати до конденсованого стану речовини. Для цього усереднення окремих
кластерів по станах ідеального газу замінюється усередненням по станах
некорельованого кристала, використовуючи
	коміркові одночастинкові потенціали. В результаті отримано розвинення
статистичної суми по кореляціях на базисі одночастинкових функцій, які
відповідають мультиплікативному наближенню. Ґраткові потенціали, що
визначають вказані функції, знаходяться з умови мінімізації залишку в
сконструйованому розвиненні.
	\keywords ґраткові моделі, групові розвинення, одночастинковий ґратковий потенціал, вільна енергія
	
\end{abstract}

\end{document}